\documentclass[final]{svjour2}
\usepackage{graphicx}
\usepackage{rotating}
\usepackage{amssymb}
\usepackage{mathptmx}
\usepackage[numbers]{natbib}
\usepackage{mathrsfs}
\usepackage{epsf}
\usepackage{psfrag}
\usepackage{dcolumn}
\usepackage{bm}
\usepackage{setspace}
\usepackage[geometry]{ifsym}
\usepackage[dvips]{color}
\usepackage{subfigure}
\makeatletter
\journalname{Journal of Low Temperature Physics}

\bibpunct{}{}{,}{s}{}{,}

\begin{document}

\newcommand{\hdblarrow}{H\makebox[0.9ex][l]{$\downdownarrows$}-}
\title{Towards Ultra-Low-Noise MoAu Transition Edge Sensors}
 
\author{D. J. Goldie$^1$ \and  A. V. Velichko$^1$ \and D. M. Glowacka $^1$ \and S. Withington$^1$}

\institute{1:Detector and Optical Physics Group, Cavendish Laboratory,\\ University of Cambridge,
J. J. Thomson Avenue, Cambridge, CB3 0HE, UK\\
Tel.:+1223 337366\\
\email{d.j.goldie@mrao.cam.ac.uk}
}

\date{24.05.2011}

\maketitle
\begin{abstract}

We report initial measurements on our first MoAu Transition Edge Sensors (TESs). The TESs
formed from a bilayer of 40 nm of Mo and 106 nm of Au showed  transition temperatures of about
320~mK, higher than identical TESs with a MoCu bilayer which is consistent with a reduced
electron transmission coefficient between the bilayer films. We report
measurements of thermal conductance in the 200~nm thick silicon nitride ${\rm SiN}_x$  
support structures at this temperature, TES dynamic behaviour and current noise measurements. 

\end{abstract}

\section{Introduction}
\label{sec:Intro}

Transition edge sensors (TESs) have become the detectors of choice
for current and future ground and space-based astronomical instruments. 
For example, the far infrared instrument SAFARI that will operate on the
joint JAXA/ESA mission SPICA 
requires state-of-the-art TESs with phonon-noise limited 
noise equivalent powers (NEPs) and fast response times.\cite{Bruce_SPICA} 
We recently reported our first measurements on ultra-low-noise MoCu TESs
with saturation powers, response times and dark NEPs close-to SAFARI requirements.\cite{Goldie_trp1} 
These MoCu TESs have reached a high degree of sophistication in terms of their
reliability in processing and reproducibility of characteristics. 
There are draw-backs, however, not least the complex chemistry of
Cu etching and the necessity for a passivating ${\rm SiO_2}$ layer
to prevent Cu corrosion. The additional heat capacity of the passivation 
increases response times and contributes additional noise.
MoAu TESs are expected to have comparable performance
and have already been demonstrated by a number of groups. 
Here we explore MoAu TESs fabricated on very thin silicon nitride
(${\rm SiN_x}$) suitable for ultra-low-noise detectors. 
These are the  first measurements of MoAu TESs fabricated 
by the Detector Physics Group in Cambridge. 
\section{Experimental details}
\label{sec:Experimental Details}
\begin{figure}[ht]
\centering

\subfigure[ $\qquad\qquad\qquad\qquad\qquad\qquad$]{
   \includegraphics[height=4.5 cm, angle=0]{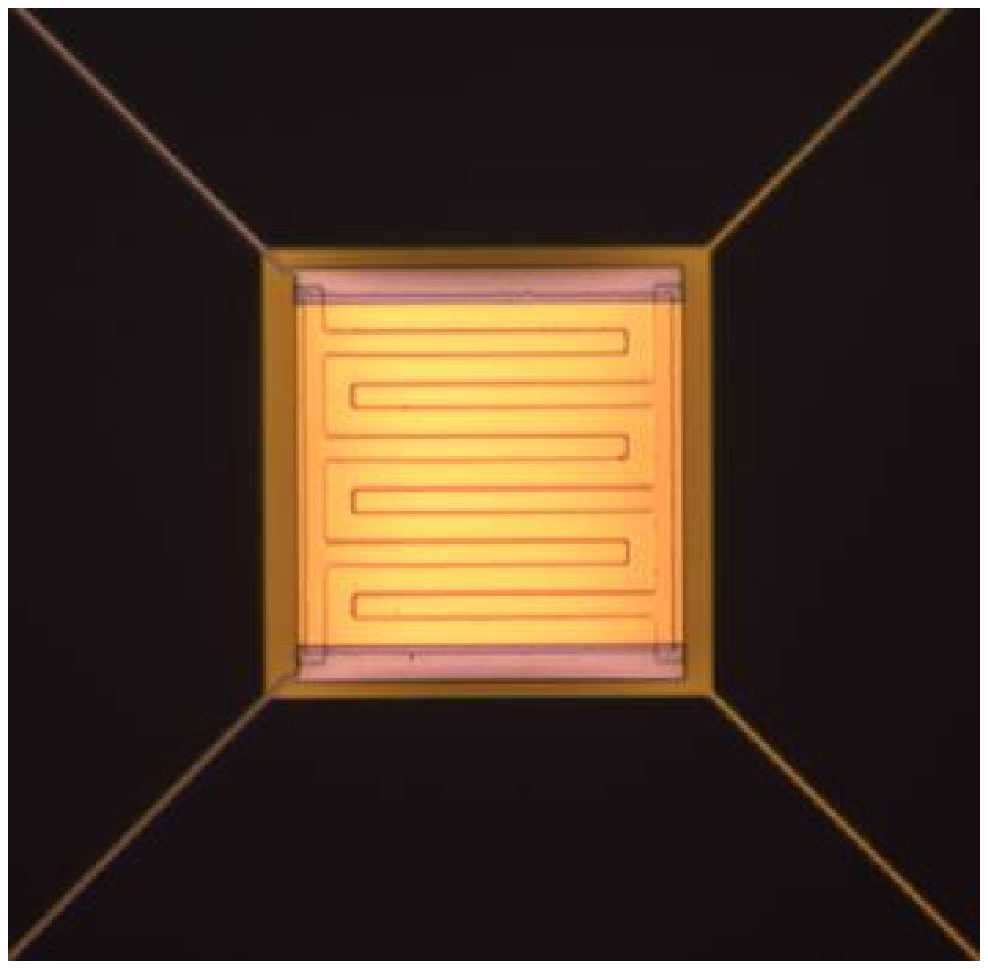}
\label{fig:fig0}

}
\subfigure[ $\qquad\qquad\qquad\qquad\qquad\qquad\qquad$]{
   \includegraphics[height=4.5 cm, angle=0]{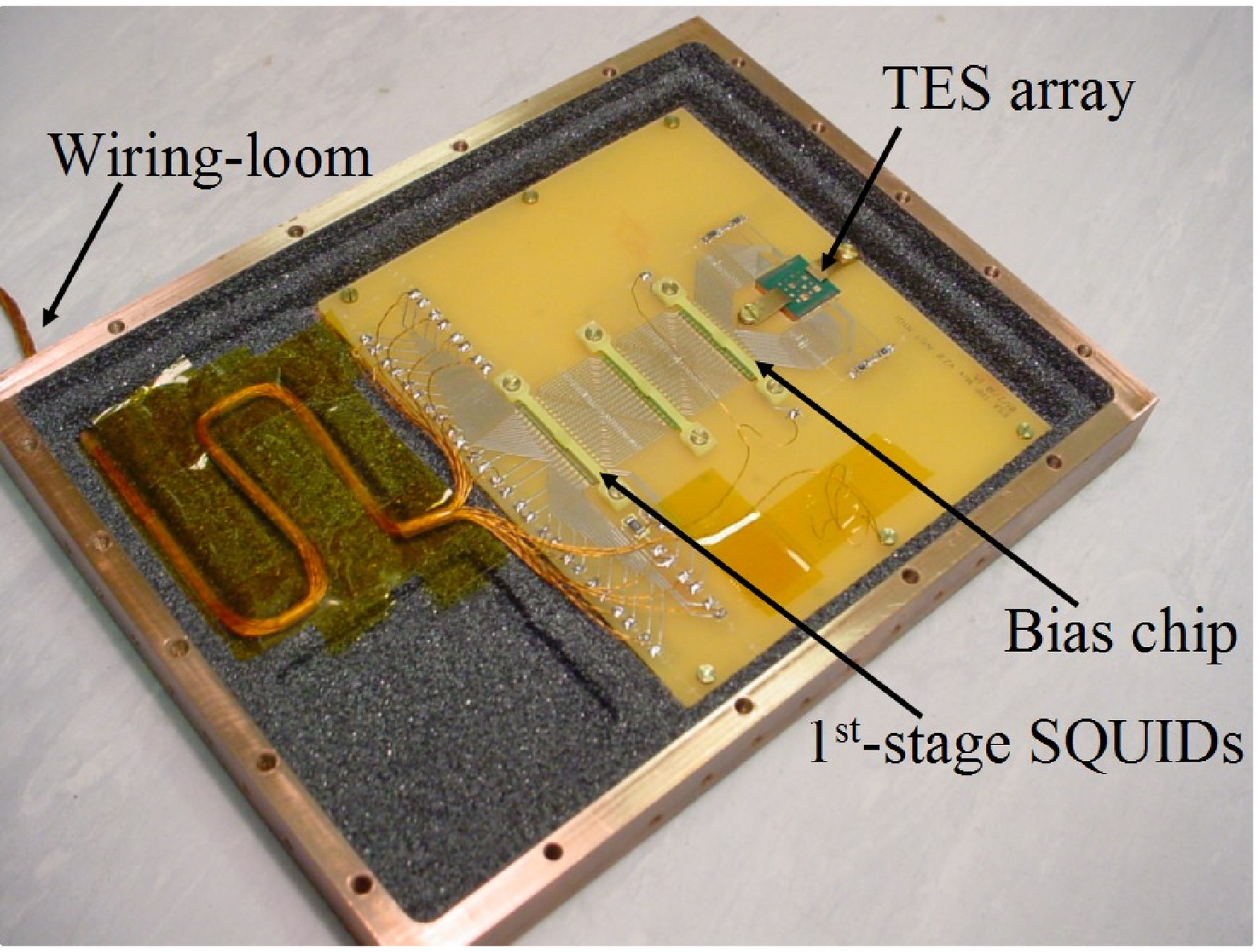}
\label{fig:blackbox}
}
\label{fig:Experimental_pictures}
\caption[Optional caption for list of figures]{(Color online) \subref{fig:fig0}  
 A single Mo/Au TES with longitudinal and partial lateral Au bars across the bilayer. The ${\rm SiN}_x$ island has an area
$110\times 110\, {\rm \mu m^2}$  and is 200~nm thick.
The supporting legs are $4\,{\rm \mu m}$ wide.
\subref{fig:blackbox} 
Photograph of the blackened
light-tight experimental enclosure. The TES array is in the upper corner of the box.
Wiring enters through a meandering labyrinth and a cover completes the assembly.
}
\end{figure}
%
The TESs reported in this paper consisted of a superconducting MoAu bilayer 
formed on a 200~nm-thick ${\rm SiN}_x$ island isolated from the heat bath by 4, long, narrow nitride legs
with widths of $2.1$ or $4.2\,{\rm \mu m}$ and lengths ranging from 
$220$ to $960\,{\rm \mu m}$.   
The TESs were formed from a bilayer consisting of 40~nm of Mo with 
106~nm of Au deposited under ultra-high vacuum by dc magnetron sputtering.
In contrast to  our earlier MoCu TESs a passivation layer to prevent corrosion of the 
exposed metal surface was not necessary.
Fig.~\ref{fig:fig0} shows a photograph of one of the MoAu TESs measured in this study. 
The released TES is almost completely flat and we estimate a curvature of order
100~nm into the plane of the figure due to slight stress in the Nb wiring which had
a transition temperature $T_c=9.2\,{\rm K}$.
%
 This curvature is much
reduced from that seen in our 
MoCu TESs on such thin nitride\cite{Goldie_trp1}
where the residual stress associated with the passivating ${\rm SiO_2}$
determines the flatness. 
More details of the fabrication techniques, resistivities and resistance ratios
 are given in Glowacka {\it et al}.\cite{Dorota2010}
%

%
\begin{table}[ht]
\caption{\label{table:fitparameters} Measured conductances and derived $T_c$'s.}
\begin{center}
\begin{tabular}{|@{}c@{}|c|c|c|c|c|}
\hline
\rule[-1ex]{0pt}{3.5ex} TES\# & $w\times L$ & $n$       &  $K_b$ & $T_c$ & $G_b$  \\
\hline
\rule[-1ex]{0pt}{3.5ex}      &      &             &  ${\rm pW/K^n }$ & ${\rm mK}$  & ${\rm pW/K} $  \\
\hline
\rule[-1ex]{0pt}{3.5ex}  1,1 &   $2.1\times420$ & 2.0	 & 1.4     & 323   &  0.88 \\
\hline
\rule[-1ex]{0pt}{3.5ex}  1,2 &  $4.2\times420$  &1.95	 & 2.7     & 321   &  1.78   \\
\hline
\rule[-1ex]{0pt}{3.5ex}  1,3 &  $2.1\times240$  &2.0	 & 1.7     & 319   &  1.1   \\
\hline
\rule[-1ex]{0pt}{3.5ex}  1,4 &  $4.2\times320$  &2.1	 & 3.2     & 317   &  2.07   \\
\hline
\rule[-1ex]{0pt}{3.5ex}  2,1 &  $4.2\times960$  &1.8	 & 1.7     & 321   &  1.24   \\
\hline
\rule[-1ex]{0pt}{3.5ex}  2,2 &  $2.1\times260$  &1.9	 & 1.7     & 321   &  1.17   \\
\hline
\rule[-1ex]{0pt}{3.5ex}  2,3 &  $4.2\times740$  &1.75	 & 1.6     & 316   &  1.18   \\
\hline
\rule[-1ex]{0pt}{3.5ex}  2,4 &  $2.1\times160$  &2.0	 & 2.1     & 315   &  1.34   \\
\hline
\rule[-1ex]{0pt}{3.5ex}  3,1 &  $2.1\times380$  &1.8	 & 1.6     & 319   &  1.17   \\
\hline
\rule[-1ex]{0pt}{3.5ex}  3,2 &  $4.2\times540$  &1.8	 & 2.5     & 316   &  1.81   \\
\hline
\rule[-1ex]{0pt}{3.5ex}  3,3 &  $2.1\times220$  &1.7	 & 1.8     & 315   &  1.35   \\
\hline
\rule[-1ex]{0pt}{3.5ex}  3,4 &  $4.2\times380$  &1.8	 & 2.2     & 312   &  1.55   \\
\hline
\rule[-1ex]{0pt}{3.5ex}  4,1 &  $4.2\times840$  &1.8	 & 1.9     & 314   &  1.39   \\
\hline
\rule[-1ex]{0pt}{3.5ex}  4,2 &  $2.1\times320$  &2.1	 & 2.4     & 313   &  1.46   \\
\hline
\rule[-1ex]{0pt}{3.5ex}  4,3 &  $4.2\times640$  &1.8	 & 1.8     & 312   &  1.29   \\
\hline
\end{tabular}
\end{center}
\end{table}

The TESs were cooled in a closed-cycle, sorption-pumped dilution refrigerator 
mounted on a pulse-tube cooler giving a base temperature of 68~mK.\cite{Teleberg2008} 
The chip, which had 16 individual TESs formed in a $4\times4$ square array, was enclosed in 
a Au-plated Cu box the inside of which is coated with light-absorbing
SiC granules and carbon black mixed in Stycast 2850 to minimize scattered light. 
A photograph of the experimental enclosure is shown in 
Fig.~\ref{fig:blackbox}.
The sample space
was surrounded by multiple layers of Nb foil and Metglas to provide
magnetic shielding. We used a NIST SQUID multiplexer\cite{DeKorteMUX} with
analog electronics readout that keeps 
the multiplexer in a fixed state but
none-the-less permits readout of multiple channels.
We identify the individual TESs by their
row and column numbers (r,c). 
\section{Results}
\label{sec:Results}
\subsection{Conductance measurements}
\label{subsec:Conductance_measure}
\begin{figure}[ht]
\centering
\subfigure[$\qquad\qquad\qquad\qquad\qquad\qquad$  ]{
\psfrag{xaxis}[] [] {$V_{TES}$ (nV)}
	\psfrag{yaxis}[] []{$P_J$ (pW)}
   \includegraphics[height=5.6 cm, angle=-90]{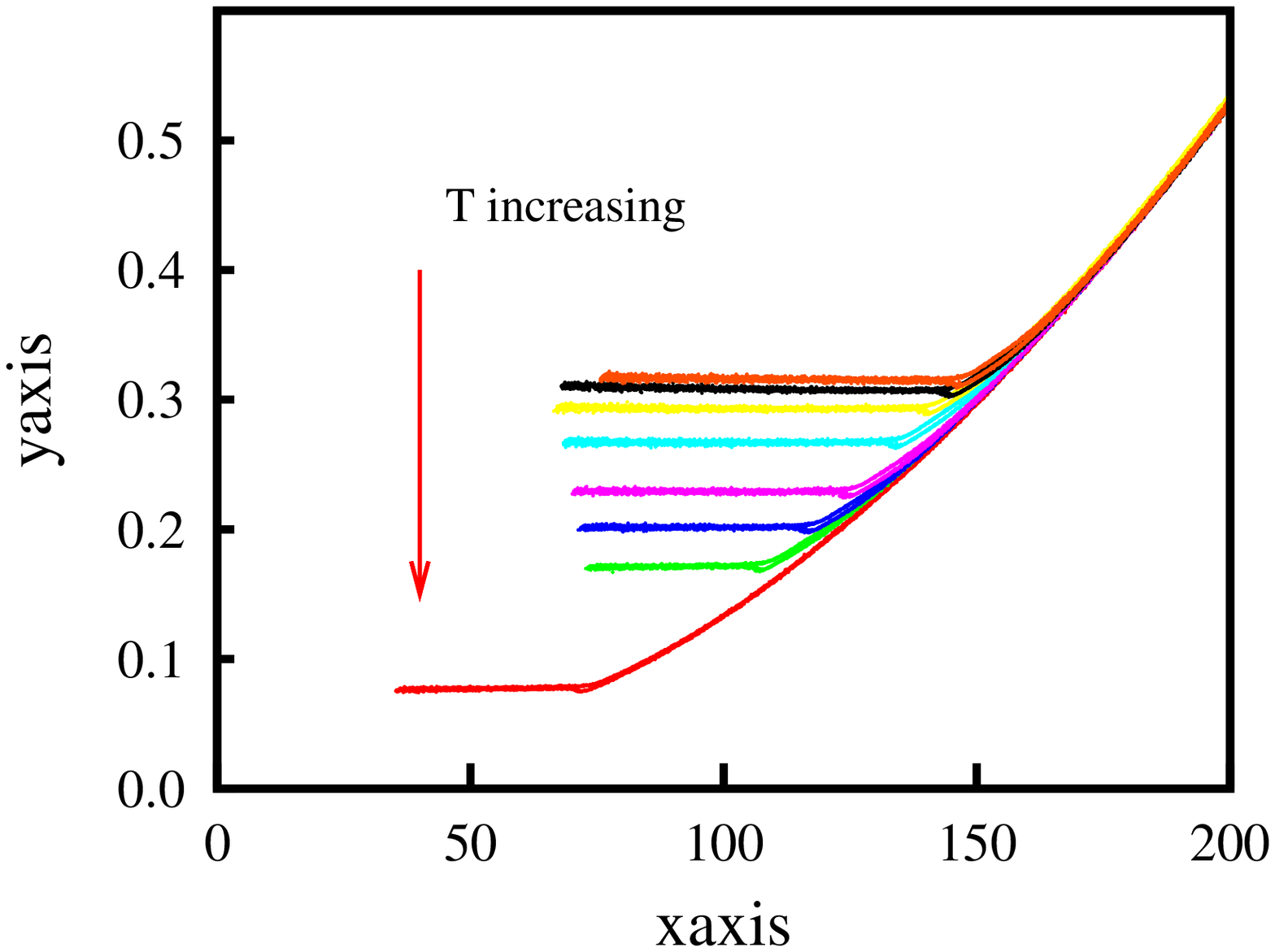}
\label{fig:power_plot}
}
\subfigure[$\qquad\qquad\qquad\qquad\qquad\qquad$ ]{
\psfrag{X}[] [] {TES No.}
	\psfrag{T}[] []{$T_c$ (mK)}
\includegraphics[height=5.6 cm, angle=-90]{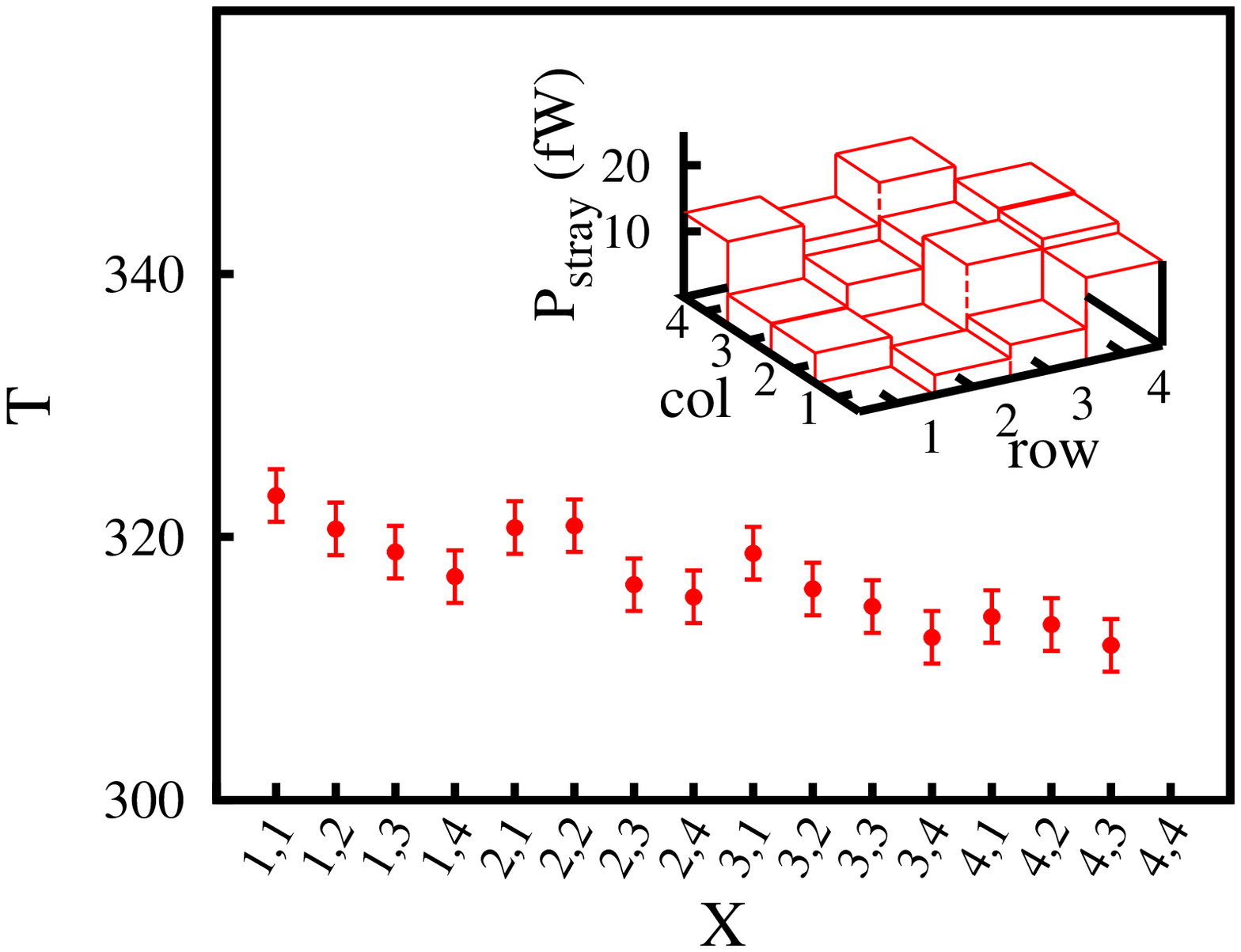}
\label{fig:T_c_variation}
}
\label{fig:Experimental_results}
\caption[Optional caption for list of figures]{
(Color online) \subref{fig:power_plot} 
Measured Joule power $P_J$ as a function of TES bias voltage $V_{TES}$
for bath temperatures of $81\,{\rm mK}$, (upper-most curve) reducing for bath temperatures of
97, 121, 148, 179, 210, 224 and 290~mK (lowest trace). 
\subref{fig:T_c_variation} 
Variation of measured $T_c$ as a function of array identification (Row,Column). The inset shows the 
estimated level of stray power across the array if the highest $T_c$ is assumed to be unaffected by stray light.
}
\end{figure}
Figure~\ref{fig:power_plot} shows the Joule power dissipation of TES~(1,4) 
for bath temperatures $T_b$ in the range 81 to 290~mK.
 The flatness of the 
power plateaux down to $0.2R_n$ are indicative of high values of the temperature-resistance coefficient $\alpha_I$
and  reflects the smoothness of the R(T) transition. 
Power dissipation as a function of $T_b$ for 15 TESs on the chip was measured. The power-flow
was modelled as $P_J=K_b(T_c^n-T_b^n)$ where $P_J$ is the Joule power which was used
to fit the  measurements with
$K_b$, $n$ and $T_c$ as free parameters. Conductances to the heat bath
$G_b=nK_bT_c^{(n-1)}$ were then calculated. Results are shown in Table~1. $T_c$
determined in this way agreed to within $2\,{\rm mK}$ with observations of the onset of supercurrent. 
The mean value of the exponent is $\overline{n}=1.9\pm 0.1$. For our higher-$G_b$ MoCu TESs
we found $\overline{n}\sim 3$ using 500~nm thick ${\rm SiN_x}$ for
nitride widths down 
to $10\,{\rm \mu m}$, lengths in the range 40 to $10\,{\rm \mu m}$ with $T_c$ in the range of 200 to 400~mK. 
For our ultra-low-$G_b$ TESs on 200~nm thick nitride and $T_c\sim 120\,{\rm mK}$ we found
$n$ in the range $1.1$ to $2.4$ and $\overline{n}= 1.8\pm 0.3$. 
This change in $n$ as a function of nitride thickness and measurement
temperature  
is characteristic of a change in  dimensionality of the heat transport. 
Normal state resistance $R_n$ was in the range $78\pm3\,{\rm m\Omega}$ for all of the TESs
measured here.
\subsection{${\bf T_c}$  variation}
\label{subsec:Tc_var}
%
Figure~\ref{fig:T_c_variation}
 shows the  derived, apparent $T_c$'s
 of 15 TESs on the measured chip plotted as a function of TES identification number. 
 The variation in $T_c$ is greater than we measured in 
our earlier high-$G_b$ MoCu TESs (which had $G_b\sim 100\,{\rm pW/K}$), and shows 
a pattern of variation strikingly similar to that observed in our 
very low-$G_b$ MoCu TESs (with $G_b\sim 0.2\,{\rm pW/K}$),  measured in the same 
cryostat and experimental set-up.\cite{Goldie_trp1} The variation and pattern is independent of the chip position on the 
wafer. 
The spread of $T_c$'s ($\pm 3.5\,{\rm mK}$) is less than observed at the lower temperature.
Note how the TESs in column~4 seem to show the greatest reduction in $T_c$ from the mean. 
Previously we thought that the variation in $T_c$ might be due to low levels of stray light
and estimated a loading of order 2 to 4~fW. In the inset of 
the figure we plot estimates of the stray light power $P_{stray}$  as a function of the position of the TES
in the array and we have assumed that the highest transition temperature $T_c^{max}$ is unaffected by stray light
so that $P_{stray}=G_b(T_c^{max}-T_c)$. Strictly this means that $P_{stray}$ is the differential loading.  
The pattern of stray light is suggestive of power incident on the far corner of the array (i.e. Row 4 and Column 4)
viewed from the perspective of the figure.

\subsection{Dynamics, Noise and Modelling}
\label{sec:Modelling}

\begin{figure}[ht]
\centering
\subfigure[$\qquad\qquad\qquad\qquad\qquad\qquad$]{
\psfrag{ylabel}[] [] {$\delta I$ $({\rm \mu A}) $ }
	\psfrag{xlabel}[] []{ Time (ms) }
	\psfrag{ylabel2}[] [] {${\rm Im}\, (Z)$ (${\rm m\Omega}$) } 
	\psfrag{xlabel2}[] [] {${\rm Re}\, (Z)$ (${\rm m\Omega}$) } 
   \includegraphics[height=5.5cm, angle=-90]{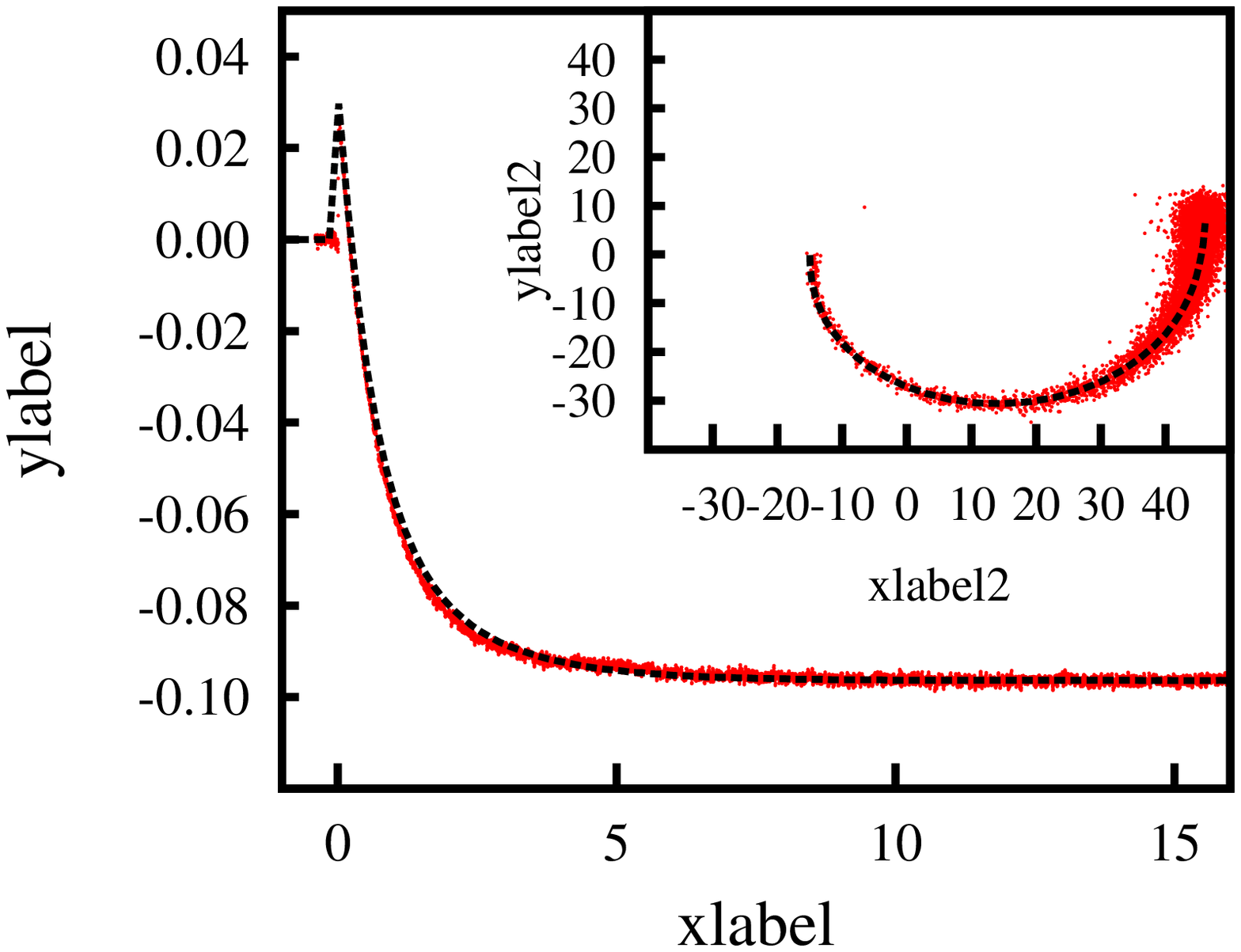}
 \label{fig:impedance_and_risetime}
}
\subfigure[$\qquad\qquad\qquad\qquad\qquad\qquad$]{
\psfrag{f}[] [] {f (Hz)}
	\psfrag{Inoise}[] []{ $ I_n \, {\rm (pA/\sqrt{Hz}})  $ }
   \includegraphics[height=5.5cm, angle=-90]{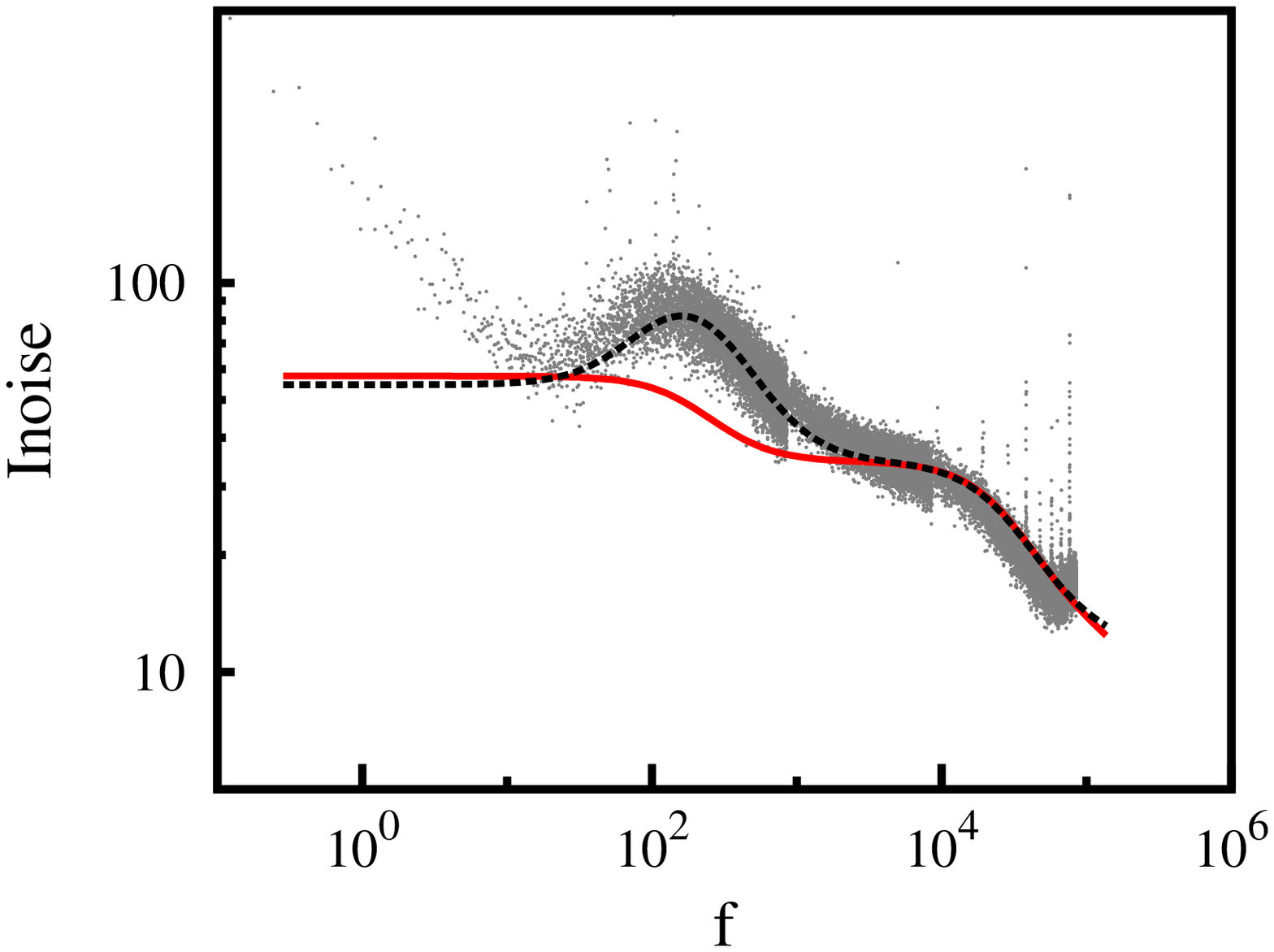}
  \label{fig:current_noise}
}
\label{fig:subfigureExample}
\caption[Optional caption for list of figures]{
(Color online) \subref{fig:impedance_and_risetime} Measured current response  to a step change in the bias voltage
and (black dashed line) the calculated response for TES(1,4) biased at $0.25R_n$. The inset
shows the real and imaginary components of the impedance and (black dashed line) 
calculation using the same parameter set. 
The  calculations use an extended thermal model for the TES with an 
additional heat capacity  of $15\, {\rm fJ/K}$ loosely coupled to the TES.
\subref{fig:current_noise} Measured current noise for the same TES at the same bias.
The solid (red) curve 
is calculated with the expected heat capacities, the dashed (black) curve
has the  additional $15\, {\rm fJ/K}$ heat capacity used to model \subref{fig:impedance_and_risetime}. 
In both cases $\gamma_{\phi}=1$.  
}
\end{figure}

 The
TES impedance $Z(f)$, the current response to a small step change in the bias voltage $\delta I$ and 
noise were measured for TES (1,4) as a function of bias. 
Figure~\ref{fig:current_noise}
shows results with $R_0=0.25R_n$
which gave the fastest observed risetime. 
A distributed thermal model to account for the expected heat capacities and conductances, including the 
${\rm SiN}_x$\cite{Karwan_SPIE},
was used. 
Figure~\ref{fig:impedance_and_risetime}
shows both the measured risetime and (inset) the impedance and calculations with an additional heat capacity 
of $C_{ex}=15\,{\rm fJ/K}$ coupled to the TES 
with  $G_{ex}= 12.5\,{\rm pW/K}$. The excess is of unknown origin. 
For these calculations $\alpha_I=380$, and the temperature-current
sensitivity $\beta_I=1.4$. 
The account of both the risetimes and impedance is very good using 
identical parameter sets and as expected $\alpha_I$ is large.

Figure~\ref{fig:current_noise} shows  measured and calculated 
current noise spectra. At frequencies above the
read-out $1/f$ knee, the measured spectrum is very-well
accounted for by the same extended thermal model 
used to describe $\delta I$ and  $Z(f)$. The phonon noise modifier
is here set to $\gamma_{\phi}=1$. This is an {\it  upper-limit} on the noise modifier
if the measurements are affected by stray light. 
Calculation of the possible contribution to the noise from stray light
is difficult since the spectrum, and the absorption and coupling efficiencies are not known.
However, spectrally, noise from
 stray light would contribute to the current noise just as phonon noise. 

\section{Summary and Conclusions}
\label{sec:Summary}

We have reported our first measurements on MoAu bilayer TESs
consisting of 40~nm Mo, 106~nm Au. 
Conductances to the heat bath have been determined for 15 TESs
each having four support legs, with widths of $2.1$ or $4.2\,{\rm \mu m}$ and lengths of
$220$ to $960\,{\rm \mu m}$ on 
200~nm thick ${\rm SiN_x}$ and are in the range 0.88 to $2.1\,{\rm pW/K}$. 
For all geometries the exponent in the power-flow is $n=1.9\pm 0.1$ which
is comparable with  identical geometries measured at $120\,{\rm mK}$ where
we found $n= 1.8\pm 0.3 $. Both values are  lower than those found for thicker nitride
films of comparable dimensions measured with transition temperatures between 200 and $400\,{\rm mK}$
where we found $n\sim 3$. 

Transition temperatures derived from 
measurements of the Joule power were
$T_c=317 \pm 3.5\,{\rm mK}$ and normal state resistances $78\pm3\,{\rm m\Omega}$. 
The variation in derived $T_c$'s show similar 
geometric patterns to earlier
measurements on very low-$G_b$ TESs measured in the same experimental
arrangement. 
This observation may be consistent with the presence of very low 
levels of stray light despite the 
use of custom-designed light shielding, held at the
cryostat base temperature, to avoid radiation from higher temperature 
sources. Estimates of the levels of differential 
stray light give $P_{stray} =8.9\pm 5\,{\rm fW}$ about a factor 4
higher than the earlier measurements. 
This emphasises the great care required when measuring low-$G_b$ structures and
the strict requirements on stray-light control on an instrument.

A good account of the dynamic behaviour is obtained using an extended thermal model for the
TES with expected heat capacities and conductances plus a small additional heat capacity of unknown origin
that is loosely coupled to the TES. 
The current noise is consistent with a phonon-noise modifier
$\gamma_{\phi}=1$ but this is
 an upper limit due to the possible influence of stray light.
In future work we will iterate the Au thickness to achieve the lower $T_c$'s required
for ultra-low-noise operation and also will include the thin-film absorbing
structures required to fabricate complete infrared detectors. Results will be published separately.

\section{Acknowledgments }
This work was supported in part by  ESA TRP
Contract No. 22359/09/NL/CP. We are  also very grateful  to colleagues working on that contract 
within the Astronomy Instrumentation Group, Cardiff
University, the
Space Research Organization of the Netherlands, the National University of Ireland, Maynooth 
and the Space Science and Technology Department of Rutherford Appleton Laboratory for numerous stimulating discussions.


\end{document}